\documentstyle[12pt]{article}

\begin{document}

\begin{titlepage}

\begin{flushright}
HUPD-9521 \\
KEK Preprint 95-109 \\
October, 1995 \\
\vspace{0.5cm}
\end{flushright}

\begin{center}
   \LARGE
    Lattice Heavy Quark Effective Theory \\
    and the Isgur-Wise function \\[1cm]

{\large Shoji Hashimoto} \\
    \normalsize
    \it National Laboratory for High Energy Physics (KEK), \\
    \it Tsukuba 305, Japan \\[3mm]
    and \\[3mm]
    {\large Hideo Matsufuru} \\
    \normalsize
    \it Department of Physics, Hiroshima University, \\
    \it Higashi-Hiroshima 739, Japan
\end{center}
\vspace{2.5cm}
\begin{abstract}
We compute the Isgur-Wise function
using heavy quark effective theory formulated on the lattice.
The non-relativistic kinetic energy term of the heavy quark
is included to the action as well as terms remaining in the
infinite quark mass limit.
The classical velocity of the heavy quark is renormalized
on the lattice and we determine the renormalized velocity
non-perturbatively using the energy-momentum dispersion
relation.
The slope parameter of the Isgur-Wise function at zero recoil
is obtained at $\beta=6.0$ on a $24^3\times 48$ lattice
for three values of $m_{Q}$.
\end{abstract}
\end{titlepage}

\newpage
\normalsize

\section{Introduction}

In the determination of the CKM (Cabibbo-Kobayashi-Maskawa)
matrix element $|V_{cb}|$ from experiment of the
exclusive decay $B\rightarrow D^{(*)}l \bar{\nu}$
the heavy quark symmetry plays an essential role,
because the universal form factor $\xi(v\cdot v')$, so
called the Isgur-Wise function, is normalized to be one at
zero recoil in the infinite heavy quark mass limit.
However the differential decay width for this process
disappear at the zero recoil point, so that one needs to
extrapolate the experimental data to this point.
The lattice computation of the slope $\xi'(1)$ enables this
extrapolation in a model independent way.
On the lattice Bernard, Shen and Soni \cite{BSS93} and
the UKQCD Collaboration \cite{UKQCD94} have computed the
Isgur-Wise function using the propagating quarks (Wilson
fermion and clover fermion respectively) for the heavy
quark.
Since the heavy quark mass should be below the lattice
cutoff in this framework, present simulations
are performed around the charm quark mass.
An alternative way of treating the heavy quark is to use
the heavy quark effective theory (HQET) on the lattice.
Mandula and Ogilvie formulated the Lattice HQET
and applied it to a calculation of the Isgur-Wise
function\cite{MO92}.
Their simulation, however, suffers from much noise
so that the extraction of the ground state seems to be
difficult.
This problem is well known in the calculation of the
heavy-light decay constant using the static approximation
where one is forced to use some method to enhance the signal
of the ground state\cite{review93} or to introduce
$O(1/m_{Q})$ terms in order to decrease the statistical
noise\cite{Hashimoto94}.

Our calculation is based on the lattice HQET keeping a part
of the $O(1/m_{Q})$ correction terms.
The inclusion of the kinetic term of the heavy quark reduces
the statistical noise significantly
and enables us to extract the ground state reliably.
We calculate the Isgur-Wise function for three values of the
heavy quark mass and discuss $O(1/m_{Q})$ effect on it.

The plan of the paper is as follows.
Section 2 contains the formulation of the lattice HQET including
$O(1/m_{Q})$ corrections.
In Section 3 we describe our numerical simulations and
discuss the results.
In this section, after brief account of simulation parameters,
we first mention about the renormalization of classical velocity.
Then the calculation of the Isgur-Wise function is described.
Sec. 4 is devoted to the conclusion.

\section{HQET on the lattice}

The action of the lattice HQET
including $O(1/m_{Q})$ corrections is
\begin{equation}
  S_{h} = \sum_{x} \bar{h}(x)\left[
           -i v{\cdot}D - \frac{1}{2m_{Q}}
            (D^{2}+\frac{1}{2}\sigma_{\mu\nu} F_{\mu\nu})
                    \right] h(x)
\label{action} \\
\end{equation}
where $v_{\mu}=(\vec{v},iv^{0})$ is a four-velocity of the
moving heavy quark.
$D_{\mu}$ is a lattice covariant derivative
and $F_{\mu \nu}$ is a chromo-magnetic (electric) field.
This action is a generalization of the non-relativistic
lattice QCD (NRQCD) for finite velocities.
The first term corresponds to the action in the infinite mass
limit used by Mandula and Ogilvie \cite{MO92}
and the second term describes the $O(1/m_{Q})$ corrections.
The heavy quark field $Q(x)$ is expressed in terms of $h(x)$
as
\begin{equation}
Q(x) =  e^{i m_{Q}v\cdot x}
    \left[ 1-\frac{1}{2m_{Q}}\not{\!\!D_{\perp}}\right]h(x),
\label{field} \\
\end{equation}
where $D_{\perp\mu} = D_{\mu}+v_{\mu}( v \cdot D )$.
For simplicity we neglect the spin dependent interaction
term $\frac{1}{4m_{Q}} \sigma_{\mu \nu} F_{\mu \nu}$ from the
action and the $O(1/m_{Q})$ correction term from the field
redefinition, i.e. the second term in eq. (\ref{field}).
It should be noted that the $O(1/m_{Q})$ corrections are not
included completely with this approximation,
but the kinetic term describes the motion of the heavy quark
inside the meson and gives a major $O(1/m_{Q})$ effect.

The heavy quark propagator is obtained by solving an
evolution equation
\begin{eqnarray}
G(x+\hat{t})
  & = & U_{4}^{\dag}(x) \left(
                        1-\frac{1}{n}H
                        \right)^{n} G(x) \nonumber \\
  &  &   + \mbox{( source term )}
\end{eqnarray}
and
\begin{equation}
H = \frac{1}{v^{0}}\left\{
      - i \vec{v}\cdot\vec{D} - \frac{1}{2m_{Q}}\left(
        \frac{-1}{(v^{0})^{2}}(\vec{v}\cdot\vec{D})^{2}
        + \vec{D}^{2} \right )
        \right\}
\end{equation}
where we used the equation of motion in the infinite mass
limit
\begin{equation}
  (  v^{0} D_{4} - i \vec{v}\cdot\vec{D} ) h(x) = 0
\end{equation}
to remove the $D_{4}^{ 2}$ term from the evolution equation.
$D_{i}h(x)$ and $D_{i}^{\,2}h(x)$ are defined as
\begin{eqnarray}
D_{i}h(x) &=& \frac{ U_{i}(x)h(x+\hat{i})
               -U_{i}^{\dag}(x-\hat{i})h(x-\hat{i})}{2} ,
 \nonumber  \\
D_{i}^{\,2}h(x) &=&  U_{i}(x)h(x+\hat{i})
            +U_{i}^{\dag}(x-\hat{i})h(x-\hat{i}) -2 h(x).
\label{deriv}
\end{eqnarray}
The cost for the computation of this deterministic equation
is much less than one of obtaining the propagator for the
Wilson fermion using some iterative solver.
A parameter $n$ is introduced to stabilize
unphysical high frequency modes.
Considering the evolution equation for free field,
the stability condition to assure the convergence of the
equation is
\begin{equation}
|\; 1-\frac{1}{n}H_{0}(\vec{k})\;|\; <\; 1
    \;\;\;\;\;(\;\; \mbox{for any}\; \vec{k}\;\;),
\end{equation}
\begin{equation}
H_{0}(k) = \frac{1}{v^{0}}\left\{
         \vec{v}{\cdot}\vec{\tilde{k}}
         +\frac{1}{2m_{Q}} \left( \vec{\hat{k}}^{2}
            - (\frac{\vec{v}}{v^{0}} \cdot \vec{\hat{k}} )^{2}
                      \right)  \right\},
\label{stability}
\end{equation}
where $\tilde{k}_{i} = \sin k_{i}$ and
$\hat{k}_{i} = 2 \sin \frac{k_{i}}{2}$.
According to the definition of derivatives (\ref{deriv}), $\hat{k}$
in the last term of (\ref{stability}) should be $\tilde{k}$.
We, however, use $\hat{k}$ instead of $\tilde{k}$
because the effect of this term is small for small $\vec{v}$,
and for convenience in later calculation.
Similar condition is required in NRQCD \cite{TL91},
but in the present case it depends on the heavy quark velocity
as well as $m_{Q}$ and $n$.
Because of the presence of term
$\frac{1}{v^{0}} \vec{v} \cdot \vec{\tilde{k}}$
high frequency modes of residual momentum do not converge
with time evolution for a large value of velocity
unless $n$ is taken to be sufficiently large.
On the other hand the $O(a)$ effect becomes larger for
larger $n$, then much large $n$ is not preferable.
Fixing $n$ for a value of $m_{Q}$, magnitude of velocities
are restricted below a certain value.
For the values of parameter used in the present simulation,
$\sum_{i}|v_{i}|$ is restricted to be smaller than 0.25.

Multiplying the projection operator to $G(x)$,
the heavy quark propagator in 4-spinor representation is
obtained as
\begin{equation}
  S_{Q}(x) = G(x) \otimes \frac{1-i \not{\!v}}{2}.
\end{equation}

The Isgur-Wise function is extracted from the three point
correlation function
\begin{equation}
  \langle 0| B(v') V_{4}(v',v) B(v)^{\dag} |0 \rangle
\end{equation}
where $V_{4}(v',v)$ is a temporal component of the vector
current $\bar{h}_{v'}\gamma_{4} h_{v}$ and $B(v)$ is a
local interpolating field for an initial or a final meson state
$\overline{q} \gamma_{5} h_{v}$ for which we take the same
mass for both initial and final mesons.
The three point correlation function is related to the
Isgur-Wise function $\xi( v\cdot v')$ as
\begin{eqnarray}
G_{v \rightarrow v'}(t_{f},t_{s},t_{i})
   &=& \sum_{\vec{x}_{s}} \sum_{\vec{x}_{f}}
      \langle 0| B(v';x_{f}) V_{4}(v',v\,;x_{s})
      B^{\dag}(v;x_{i}) |0 \rangle        \nonumber \\
   &=& \sum_{\vec{x}_{s}} \sum_{\vec{x}_{f}}
      \langle\,Tr[ S_{q}(x_{f},x_{i})^{\dag}
      S_{h_{v'}}(x_{f},x_{s}) \gamma_{4}
      S_{h_{v}}(x_{s},x_{i} ) ]\,\rangle \nonumber \\
   &\propto&   m_{B}\,\xi(v \cdot v')(v+v')_{0} \nonumber \\
   &       &   \times e^{-E_{f}(t_{f}-t_{s})-E_{i}(t_{s}-t_{i})}
       \;\;\;\; (\mbox{for}\ \  t_{f} \gg t_{s} \gg t_{i} )
\end{eqnarray}
where $x_{f}=(\vec{x}_{f},t_{f})$, $x_{s}=(\vec{x}_{s},t_{s})$,
$x_{i}=(0,t_{i})$ and
the component proportional to $(v-v')_{0}$
is neglected since its form factor is
$O(\Lambda_{QCD}/m_{Q})$ and $(v-v')_{0}$ is small
itself in our velocity region.
Taking the following ratio of the three-point functions the
exponential factor and the renormalization factor for the
vector current cancel and we obtain
\begin{eqnarray}
R_{v,v'}(t_{f},t_{s},t_{i})
  &=&
       \frac{ G_{v \rightarrow v'}(t_{f},t_{s},t_{i})
                      G_{v' \rightarrow v}(t_{f},t_{s},t_{i}) }
            { G_{v \rightarrow v }(t_{f},t_{s},t_{i})
                     G_{v' \rightarrow v'}(t_{f},t_{s},t_{i}) }
  \nonumber \\
  &\rightarrow&
           | \xi( v{\cdot}v') |^{2}
                 \frac{ (v_{0}+v'_{0})^{2} }{ 4 v_{0} v'_{0}}.
\end{eqnarray}
We use this relation for the calculation of the Isgur-Wise
function.

\section{Simulations and results}

We used 120 gauge configurations of
$24^{3}$ $\times$ $48$ lattice at $\beta=6.0$
in the quenched approximation.
Each configurations are separated by 2,000 pseudo-heat bath
sweeps after 20,000 sweeps for thermalization.
For the light quarks we used the Wilson fermion with hopping
parameters 0.153 and 0.155, and obtained propagators under
the periodic and Diriclet boundary conditions for spatial
and temporal direction respectively.
The critical hopping parameter and the inverse lattice
spacing determined from rho meson mass are
$\kappa_{c} = 0.156986(50)$
and $a^{-1} = 2.215(64)$ GeV.

Heavy quark masses and stabilization parameters
we used are
\[
   \left(
  \begin{array}{l}
    m_{Q} \\
    \; n
 \end{array}
\right)
=  \left(
  \begin{array}{l}
    1.8  \\
    \; 3
 \end{array}
\right)
,  \left(
  \begin{array}{l}
    2.5 \\
    \; 2
 \end{array}
\right)
,  \left(
  \begin{array}{l}
    5.0  \\
    \; 1
 \end{array}
\right)
\]
where $m_{Q}=1.8$ nearly corresponds to the bottom quark mass.
As mentioned above,
heavy quark velocities should satisfy the stability condition
(\ref{stability}) which leads $\sum_{i} |v_{i}| \leq 0.25$
for our sets of $m_{Q}$ and $n$.
In Table 1 we tabulate our sets of $v$ and $v'$ used in
the simulation.

The mean-field improvement\cite{LM93} is applied when we
compute the heavy quark evolution equation.
Link variables are altered as
  $U_{\mu}(x) \rightarrow  U_{\mu}(x) / u_{0}$
where
$u_{0}=\langle\,\frac{1}{3}U_{plaq.}\,\rangle^{1/4}$.
$U_{plaq.}$ is the product of link variables along plaquette.
We used the value $u_{0} = 0.8776$ measured on our
configurations.

\subsection{Extraction of the renormalized velocity}

Before we proceed to the calculation of the Isgur-Wise
function we mention about the determination of masses and
velocities of the heavy-light mesons.
The velocity of the moving heavy-light meson $v_{R}$ could
differ from the one of the bare heavy quark $v$ for the
lattice HQET because of the violation of the space-time
$O(4)$ symmetry\cite{Agl94,MO94}.
We extract the `renormalized' velocity from the simulation results
nonperturbatively using the dispersion relation of the meson.

The heavy-light meson obeys the dispersion relation
\begin{equation}
  p^{0} =
m_{P}v^{0} + \frac{1}{v^{0}}\left\{
         \vec{v}{\cdot}\vec{k'}
         +\frac{1}{2m_{P}} \left( \vec{k'}^{2}
         - (\frac{\vec{v}}{v^{0}}{\cdot}\vec{k'} )^{2}
                         \right)  \right\}
\end{equation}
where $p^{\mu}$ is the meson momentum which is decomposed
as $p^{\mu}=m_{P}v^{\mu}+k'^{\mu}$.
$k'_{\mu}$ differs from our definition of the residual momentum
$k_{\mu}$ which is defined through $p^{\mu}-m_{Q}v^{\mu}$.
Using a naive mass relation
\begin{equation}
m_{P}=m_{Q}+E_{0}
\end{equation}
where $E_{0}$ is the binding energy, $k'^{\mu}$ is written
as  $k'^{\mu}=k^{\mu}-E_{0}v^{\mu}$.
Substituting this relation in above dispersion relation,
we obtain
 \begin{equation}
 k^{0}
   = \frac{E_{0}}{v^{0}}\left( 1
       +\frac{E_{0}}{2m_{P}}(\frac{\vec{v}}{v^{0}})^{2} \right)
      + \frac{\vec{v}}{v^{0}}\cdot\vec{k}
                \left( 1-\frac{E_{0}}{{m_{P}v^{0}}^{2}} \right)
      + \frac{1}{2m_{P}v^{0}}\left( \vec{k}^{2}
                -(\frac{\vec{v}}{v^{0}}\cdot\vec{k} )^2 \right).
\end{equation}
{}From now on, we write $E(k)$ instead of $k^{0}$,
and attach subscript `R' to the classical velocity $v$
for distinction from ones as the input parameter.

On the lattice, above dispersion relation is rewritten as
\begin{equation}
e^{-[E(k)-E(k=0)]}
 = 1 - \left\{  \frac{\vec{v_{R}}}{v^{0}_{R}}\cdot( \vec{\tilde{k}}
          -\frac{E_{0}}{m_{P}{v^{0}_{R}}^{2}} \vec{\hat{k}} )
          + \frac{1}{2m_{P}v^{0}_{R}} \left( \vec{\hat{k}}^{2}
          -(\frac{\vec{v_{R}}}{v^{0}_{R}}\cdot\vec{\hat{k}})^2
  \right) \right\},
\end{equation}
where
\begin{equation}
E(k=0) = \frac{E_{0}}{v^{0}_{R}} \left( 1 +
          \frac{E_{0}}{2m_{P}}(\frac{\vec{v_{R}}}{v^{0}_{R}})^{2}
                                              \right).
\label{bindingenergy}
\end{equation}

In Fig.~\ref{fig:rvext} we plot the asymmetry of the energy
$-[e^{-E(k)+E(0)}-e^{-E(-k)+E(0)}]/2$ from which we extract the
`renormalized' velocity $v_{R}$ using the relation
\begin{equation}
- \frac{ e^{-E(k)+E(0)}-e^{-E(-k)+E(0)}}{2}
 = \frac{v_{Ri}}{v^{0}_{R}}\cdot
   \left( \sin k_{i}
   - \frac{E_{0}}{m_{P}{v^{0}_{R}}^{2}}2\sin\frac{k_{i}}{2} \right).
\label{rv}
\end{equation}
For $E_{0}$ and $m_{P}$ in the right hand side of (\ref{rv}),
we use the values obtained at $\vec{v}=0$.
For the value of $v^{0}$, bare one is used.
These substitution should be justified by the independence
of these values on the velocity.
Obtained values of the `renormalized' velocity are
listed in Table ~\ref{tab:renormv} and
slightly larger than `bare' values $v$ , but in most cases
consistent with $v$ in present error level.
Ratios $v_{R}/v$ are almost independent on $v$ as shown in
Fig.~\ref{fig:deltav}.

The heavy-light meson masses can also
be obtained similarly using the relation
\begin{equation}
1-\frac{ e^{-E(k)+E(0)}+e^{-E(-k)+E(0)}}{2}
 = \frac{1}{2m_{P}v_{R}^{0}} \left(  \vec{\hat{k}}^{2}
      - (\frac{\vec{v}_{R}}{v_{R}^{0}}{\cdot}\vec{\hat{k}} )^{2}
           \right)
\end{equation}
which we plot in Fig.~\ref{fig:mpext}.
The obtained values for $m_{P}$ are summarized
in Table ~\ref{tab:mesonmass}.
In Table ~\ref{tab:Eb}, we also list the results for binding energy
of heavy-light meson, $E_{0}$, determined from correlation
function at zero residual momentum using equation
(\ref{bindingenergy}).
In eq. (\ref{bindingenergy}), we again use the values at
$\vec{v}=0$ for $E_{0}$ and $m_{P}$.
Both of $m_{P}$ and $E_{0}$ are almost independent
of the velocity and well satisfy the naive mass relation
$m_{P}=m_{Q}+E_{0}$.

\subsection{The Isgur-Wise function}

In the calculation of the three point function
we set $t_{i}=8$ and $t_{s}=20$ where it seems to reach
the ground state of initial meson moving with velocity $v$.
Fig.~\ref{fig:Roft} shows the behavior of
$R_{v,v'}(t_{f},t_{s},t_{i})$
as functions of $t_{f}$ for several sets of $v$ and $v'$
for $m_{Q}$=1.8 and $\kappa$=0.153.
Signal is rather clean and we observe clear plateaus beyond
$t_{f}\approx$ 24.
Then we take $t_{f}$=24-27 as a fitting interval.
For the largest mass $m_{Q}$=5.0 signal becomes so noisy and we
cannot identify a clear plateau.
Nevertheless we treat them in the same manner as $m_{Q}=1.8$
and $2.5$ with $t_{f}$=24-25.

Our results for the form factor $\xi(v_{R}\cdot v_{R}')$ at
$\kappa$=0.153 are shown in Fig.~\ref{fig:xi} for
$m_{Q}$=1.8, 2.5 and 5.0.
Error bars in the horizontal direction are coming from the
statistical uncertainty in the determination of the
renormalized velocity.
Statistical errors increase with heavy quark mass
as expected.
In our $v\cdot v'$ region the form factors have almost
linear behavior where we extract the slope of
$\xi(v_{R}\cdot v_{R}')$ at $v_{R}\cdot v_{R}'$=1, which is
usually denoted as $-\rho^{2}$, from a fit of our data to
the the form
\begin{equation}
  \xi ( v_{R}{\cdot}v_{R}' )
   = 1 - \rho^{2}(v_{R}{\cdot}v_{R}' -1 ).
\end{equation}
$\rho^{2}$ for each $m_{Q}$ and $\kappa$ are given in
Table ~\ref{tab:rho2} with the values extrapolated to the critical
hopping parameter for the light quark.
$m_{Q}$=1.8 roughly corresponds to the B meson mass where
our data should be compared with the experimental data
obtained by CLEO Collaboration\cite{CLEO94}.
\begin{eqnarray}
\rho^{2} & = & \mbox{0.70} \pm \mbox{0.17} \mbox{\ \ \ this work} \\
         & = & \mbox{0.84} \pm \mbox{0.15} \mbox{\ \ \ CLEO}
\end{eqnarray}
Our data is consistent with that of CLEO
considering the large statistical uncertainty.
We also compare our results with other lattice results
obtained using the propagating heavy quarks for which the
heavy quark mass is around the charm quark mass
\begin{eqnarray}
\rho^{2} & = & \mbox{1.24} \pm \mbox{0.26 (stat)}
                           \pm \mbox{0.33 (syst)}
               \mbox{\ \ \ \cite{BSS93}} \\
         & = & \mbox{0.9} {}^{+2}_{-3}\mbox{(stat)}
                          {}^{+4}_{-2}\mbox{(syst)}
               \mbox{\ \ \ \cite{UKQCD94}}.
\end{eqnarray}
These are again consistent with each other in their large
statistical uncertainty.

In order to see the $O(1/m_{Q})$ effect we
extrapolate our $\rho^{2}$ to the vanishing $1/m_{Q}$
limit as shown in Fig.~\ref{fig:infm},
$\rho^{2}$ becomes smaller when we approach the infinite
mass limit.
We need, of course, more statistics to quantify the size of
the $O(1/m_{Q})$ effect and to extrapolate our data to the
charm quark mass.

\section{Conclusion}
We calculated the Isgur-Wise function near the zero recoil
limit using HQET including the kinetic term of the heavy quark.
Spin-independent $O(1/m_{Q})$ effect is included in this
approximation.
The obtained value for the slope parameter $\rho^{2}$ is
consistent with the experimental value, but the statistical
error is still large.
A large amount of this statistical error is due to the large
statistical uncertainty in the determination of the
renormalized classical velocity
and it originates in extraction of binding energy
of heavy-light meson for finite residual momenta.
It is essential for reducing the statistical error to
improve the signal for finite momenta using the smearing
techinique for example.

It is important to include the $O(1/m_{Q})$ terms
completely and to quatify the size of violations of the
heavy quark symmetry in several semi-leptonic form factors.
The lattice HQET is also applicable for studies of heavy to
light semi-leptonic decay ($B \rightarrow \rho l \nu$ etc.)
which can be used for extraction of $V_{ub}$ from the
exclusive decay width in a model independent way.

\section*{Acknowledgment}

The numerical computations were carried out on Intel Paragon
XP/S at INSAM (Institute for Numerical Simulations and
Applied Mathematics) of Hiroshima University.
We would like to thank S. Hioki and O. Miyamura for useful
discussions.
S.H. was supported in part by the Grant-in-Aid of the
Ministry of Education under the contract No. 076117.


\clearpage

\begin{center}
TABLES
\end{center}

\begin{table}[h]
\begin{center}
\begin{tabular}{|l|l|}
\hline
\makebox[55mm]{$\vec{v} \;\;$ ( initial ) } &
\makebox[65mm]{$\vec{v'}\;\;$ ( final )   }                    \\
\hline \hline
 $(\;0,\;0,\;0)\;$   & $(\; 0,\; 0,\;v_{f}\;)$                 \\
        & \hspace*{\fill} $v_{f}=0.05, 0.10, 0.15, 0.20, 0.25$ \\
                 & $( \;0,\; v_{f},\;v_{f}\;)$                 \\
                 &  \hspace*{\fill}  $v_{f}=0.05, 0.10$        \\
\hline
 $(\;0,\;0,\;v_{i}\;)$
            & $(\;0,\;0,\;\pm v_{f}\;),(\;0,\;v_{f},\;0\;)$    \\
 \hspace*{\fill} $v_{i}=0.05, 0.10, 0.15, 0.20, 0.25$
  & \hspace*{\fill} $v_{f}=0.0, 0.05, 0.10, 0.15, 0.20, 0.25$  \\
        & $(\;0,\;v_{f},\;v_{f}\;),(\;0,\;v_{f},\;-v_{f}\;)$   \\
                   &  \hspace*{\fill}  $v_{f}=0.05, 0.10$      \\
\hline
 $(\;0,\;v_{i},\;v_{i}\;)$
                   & $(\;0,\;v_{f},\;0\;)(\;0,\;0,\;-v_{f}\;)$ \\
 \hspace*{\fill} $v_{i}=0.05, 0.10$
  & \hspace*{\fill} $v_{f}=0.0,0.05, 0.10, 0.15, 0.20, 0.25$   \\
    & $(\;0,\;v_{f},\;\pm v_{f}\;),(\;0,\;-v_{f},\;-v_{f}\;)$  \\
                 &  \hspace*{\fill}  $v_{f}=0.05, 0.10$        \\
\hline
\end{tabular}
\end{center}
\caption{
Initial and final heavy quark velocities used in our simulation.
When we take the ratio $R_{v,v'}(t_{f},t_{s},t_{i})$
rotationally equivalent combinations of $v$ and $v'$
are used.}
\label{tab:inputv}
\end{table}

\begin{table}[h]
\begin{center}
\begin{tabular}{|c||c|c|c|c|c|}
\hline
$\kappa=0.153$  & \multicolumn{5}{|c|}{bare velocity} \\
\hline
\makebox[10mm]{$m_{Q}$} & \makebox[17mm]{ 0.05 } &
\makebox[17mm]{ 0.10 }  & \makebox[17mm]{ 0.15 } &
\makebox[17mm]{ 0.20 }  & \makebox[17mm]{ 0.25 }   \\
\hline
1.8 & 0.054(04)& 0.105(05)& 0.157(07)& 0.209(10)& 0.263(14) \\
2.5 & 0.056(04)& 0.108(06)& 0.161(08)& 0.216(12)& 0.271(17) \\
5.0 & 0.059(05)& 0.112(08)& 0.166(12)& 0.222(20)& 0.279(31) \\
\hline
\hline
$\kappa=0.155$  & \multicolumn{5}{|c|}{bare velocity} \\
\hline
\makebox[10mm]{$m_{Q}$} & \makebox[17mm]{ 0.05 } &
\makebox[17mm]{ 0.10 }  & \makebox[17mm]{ 0.15 } &
\makebox[17mm]{ 0.20 }  & \makebox[17mm]{ 0.25 }   \\
\hline
1.8 & 0.052(06)& 0.102(07)& 0.152(09)& 0.204(12)& 0.255(16) \\
2.5 & 0.054(06)& 0.104(07)& 0.156(10)& 0.208(14)& 0.262(21) \\
5.0 & 0.055(07)& 0.107(10)& 0.158(15)& 0.210(23)& 0.261(35) \\
\hline
\end{tabular}
\end{center}
\caption{Results for renormalized velocities
for each $\kappa$ and $m_{Q}$.
These values are obtained from simulation data nonperturbatively
using the dispersion relation.   }
\label{tab:renormv}
\end{table}

\begin{table}[h]
\begin{center}
\begin{tabular}{|c||c|c|c|c|c|c|}
\hline
$\!\kappa\!=\!0.153\!$ & \multicolumn{6}{|c|}{bare velocity} \\
\hline
\makebox[10mm]{$m_{Q}$} &
\makebox[15mm]{ 0.00 }  & \makebox[15mm]{ 0.05 } &
\makebox[15mm]{ 0.10 }  & \makebox[15mm]{ 0.15 } &
\makebox[15mm]{ 0.20 }  & \makebox[15mm]{ 0.25 }   \\
\hline
1.8 & 2.30(06)& 2.28(06)& 2.26(07)& 2.24(09) & 2.21(12)& 2.17(16) \\
2.5 & 2.98(08)& 2.94(09)& 2.90(11)& 2.85(15) & 2.78(20)& 2.68(28) \\
5.0 & 5.39(21)& 5.31(24)& 5.17(33)& 4.99(51) & 4.81(82)&
    \makebox[15mm]{4.72(134)} \\
\hline
\hline
$\!\kappa\!=\!0.155\!$  & \multicolumn{6}{|c|}{bare velocity} \\
\hline
\makebox[10mm]{$m_{Q}$} &
\makebox[15mm]{ 0.00 }  & \makebox[15mm]{ 0.05 } &
\makebox[15mm]{ 0.10 }  & \makebox[15mm]{ 0.15 } &
\makebox[15mm]{ 0.20 }  & \makebox[15mm]{ 0.25 }   \\
\hline
1.8 & 2.30(07)& 2.29(08)& 2.27(09)& 2.26(12)& 2.24(15) & 2.20(21) \\
2.5 & 3.00(10)& 2.97(11)& 2.93(14)& 2.89(19)& 2.83(25) & 2.74(35) \\
5.0 & 5.45(25)& 5.40(29)& 5.28(40)& 5.16(61)&
    \makebox[15mm]{5.09(100)}&\makebox[15mm]{5.20(169)} \\
\hline
\end{tabular}
\end{center}
\caption{ The kinetic mass of heavy-light meson
for each $m_{Q}$ and $\kappa$ obtained
using the dispersion relation.
Obtained values are almost independent of velocity.  }
\label{tab:mesonmass}
\end{table}

\begin{table}[h]
\begin{center}
\begin{tabular}{|c||c|c|c|c|c|c|}
\hline
$\!\kappa\!=\!0.153\!$  & \multicolumn{6}{|c|}{bare velocity} \\
\hline
\makebox[10mm]{$m_{Q}$} &
\makebox[15mm]{ 0.00 }  & \makebox[15mm]{ 0.05 } &
\makebox[15mm]{ 0.10 }  & \makebox[15mm]{ 0.15 } &
\makebox[15mm]{ 0.20 }  & \makebox[15mm]{ 0.25 }   \\
\hline
1.8 & 0.531(04)& 0.530(04)& 0.529(04)&
                             0.528(04)& 0.527(05)& 0.525(06) \\
2.5 & 0.533(04)& 0.532(04)& 0.532(05)&
                             0.531(05)& 0.530(06)& 0.528(08) \\
5.0 & 0.533(06)& 0.534(06)& 0.535(08)&
                             0.537(10)& 0.539(13)& 0.544(19) \\
\hline
\hline
$\!\kappa\!=\!0.155\!$  & \multicolumn{6}{|c|}{bare velocity} \\
\hline
\makebox[10mm]{$m_{Q}$} &
\makebox[15mm]{ 0.00 }  & \makebox[15mm]{ 0.05 } &
\makebox[15mm]{ 0.10 }  & \makebox[15mm]{ 0.15 } &
\makebox[15mm]{ 0.20 }  & \makebox[15mm]{ 0.25 }   \\
\hline
1.8 & 0.497(05)& 0.497(05)& 0.496(05)&
                             0.495(06)& 0.493(06)& 0.491(07) \\
2.5 & 0.500(06)& 0.499(06)& 0.499(06)&
                             0.498(07)& 0.496(08)& 0.495(10) \\
5.0 & 0.500(09)& 0.502(09)& 0.503(10)&
                             0.505(12)& 0.509(17)& 0.514(25) \\
\hline
\end{tabular}
\end{center}
\caption{ The binding energy obtained from correlation functions
at zero residual momentum for each velocity, $m_{Q}$ and $\kappa$.
Obtained values are almost independent of velocity.  }
\label{tab:Eb}
\end{table}

\begin{table}[h]
\begin{center}
\begin{tabular}{|c||c|c|c|c|}
\hline
           & \multicolumn{4}{|c|}{$m_{Q}$} \\
\cline{2-5}
\makebox[24mm]{$\kappa$} & \makebox[24mm]{ 1.8 } &
\makebox[24mm]{  2.5   } & \makebox[24mm]{ 5.0 } &
\makebox[24mm]{$\infty$}\\
\hline
0.153        &  0.71(09) &  0.72(12) &  0.77(30) &  0.78(30)  \\
0.155        &  0.70(13) &  0.71(18) &  0.77(47) &  0.75(46)  \\
$\kappa_{c}$ &  0.70(17) &  0.69(24) &  0.77(66) &  0.73(67)  \\
\hline
\end{tabular}
\end{center}
\caption{
Results for $\rho^{2}\equiv-\xi'(1)$ for each $m_{Q}$ and $\kappa$.
They were obtained by fitting
$1-\rho^{2}(v_{R}{\cdot}v_{R}'-1)$ with simulation data.
Results at $m_{Q}=\infty$ are obtained from extrapolation
in $1/m_{Q}$ and the values at the critical hopping parameter
( $\kappa_{c}=0.156986(50)$ ) are obtained by linear
extrapolation in $1/\kappa$. }
\label{tab:rho2}
\end{table}

\clearpage


\begin{figure}[p]
\caption{
Extraction of renormalized velocity at $m_{Q}=1.8$
and $\kappa=0.153$.
It shows $\,-[e^{-E(k)+E(0)}-e^{-E(-k)+E(0)}]/2\,\,$ vs
$\,\,\sin(k_{z})-(E_{0}/m_{P}{v^{0}}^{2})\,2\!\sin(k_{z}/2)\,$
for $v=(0,0,0.10)$ and $(0,0,0.20)$, where
we use the obtained values at $\vec{v}=0$ for $E_{0}$ and $m_{P}$
and bare values for $v^{0}$.
Solid lines represent linear fits from which we
extract the renormalized velocities. }
\label{fig:rvext}
\end{figure}

\begin{figure}[p]
\caption{
$\delta{v}/v$ for $m_{Q}=1.8$ and $\kappa=0.153$.
$\,\delta{v}=v_{R}-v\,\,$ where the values of $v_{R}$ are listed in
Table 2.
$\delta{v}/v$ are independent of $v$ and slightly larger
than zero, but consistent with zero in present error level. }
\label{fig:deltav}
\end{figure}

\begin{figure}[p]
\caption{
$1-[e^{-E(k)+E(0)}+e^{-E(-k)+E(0)}]/2\,\,$  as a function of
$\,(2\sin(k_{z}))^{2}\,$ for $v=(0,0,0.20)$
at $m_{Q}=1.8$ and $\kappa=0.153$.
Solid line represents linear fit from which we
extract the kinetic mass of heavy-light meson. }
\label{fig:mpext}
\end{figure}

\begin{figure}[p]
\caption{
$R_{v,v'}(t_{f},t_{s},t_{i})$ as functions of $t_{f}$
for several sets of $v$ and $v'$ at
$m_{Q}=1.8$ and $\kappa=0.153$.
For all combinations, $v'$s have the same absolute values
as $v$ and opposite spatial directions to $v$.  }
\label{fig:Roft}
\end{figure}

\begin{figure}[p]
\caption{
The form factors $\xi(v_{R}{\cdot}v_{R}')$ at
$m_{Q}=1.8$ (a), $2.5$ (b) and $5.0$ (c) where $\kappa=0.153$ for
all cases.
Figures show the almost linear behavior of
$\xi(v_{R}{\cdot}v_{R}')$ in our velocity region.
Slope of these functions, $\rho^{2}$, are extracted by fitting
to the form $\,\,1-\rho^{2}(v_{R}{\cdot}v_{R}'-1)$.  }
\label{fig:xi}
\end{figure}

\begin{figure}[p]
\caption{
Extrapolation of $\rho^{2}$ to the vanishing $1/m_{Q}$
limit at $\kappa=0.153$.
The values of $\rho^{2}$ are listed in Table 5.}
\label{fig:infm}
\end{figure}

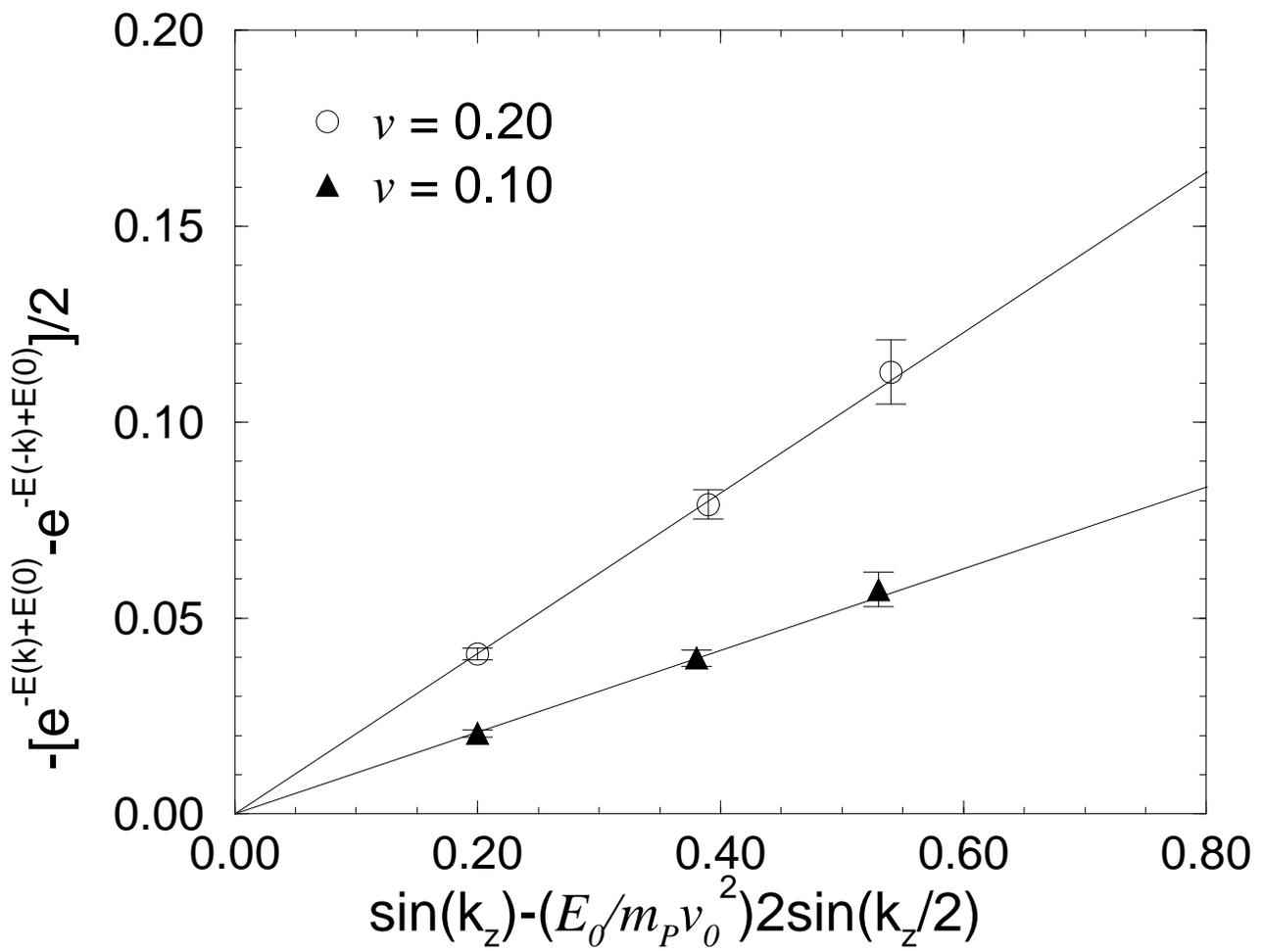
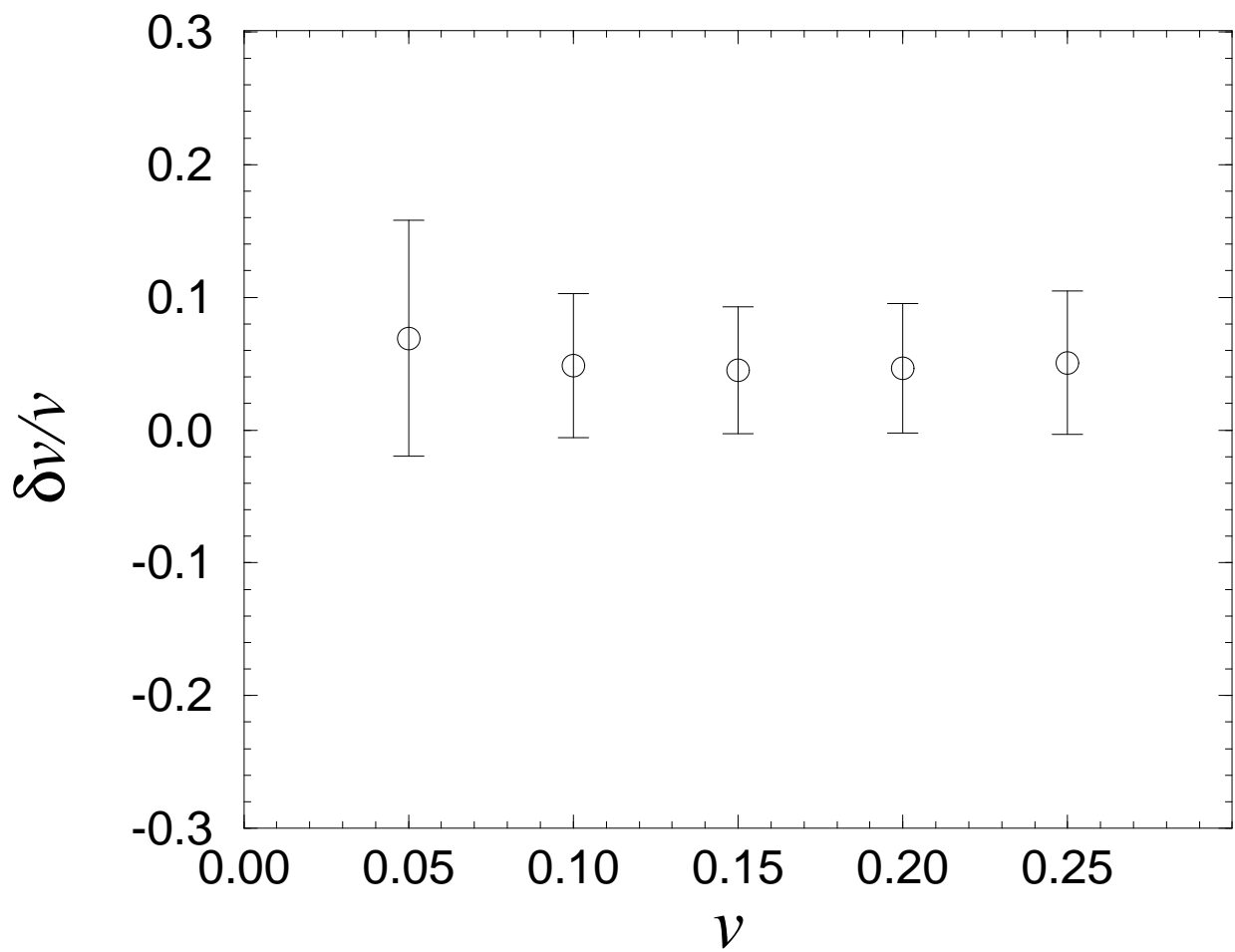
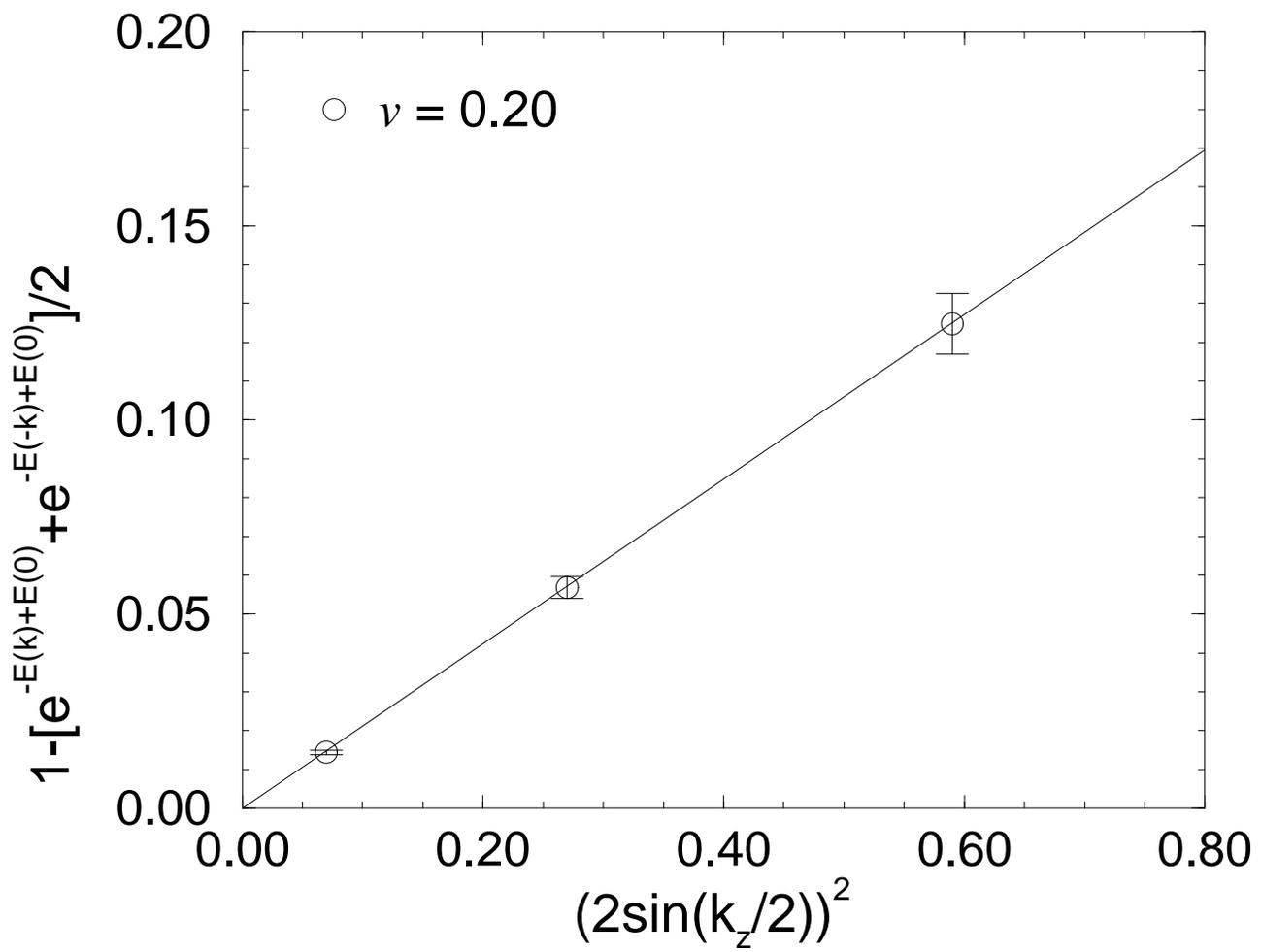
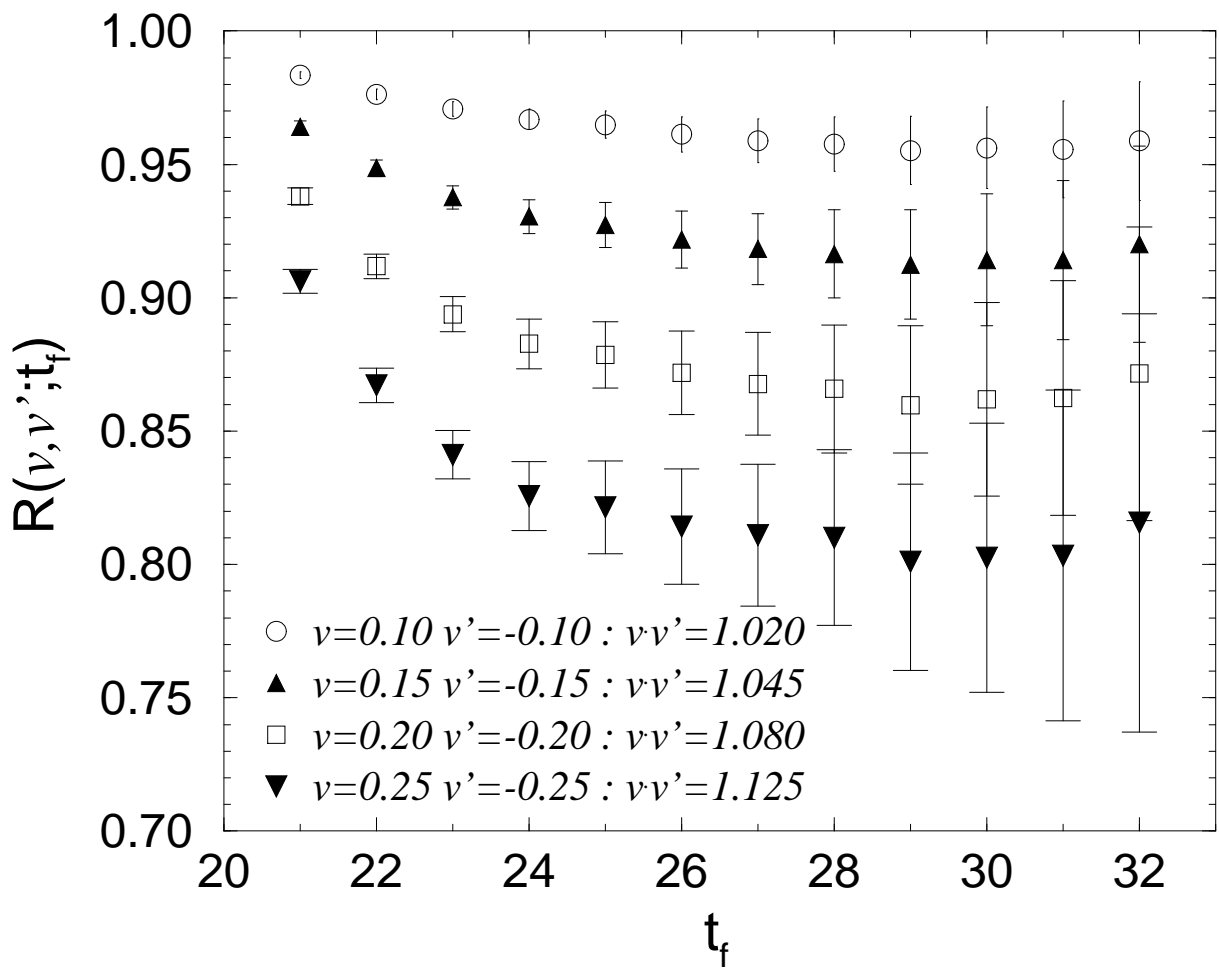
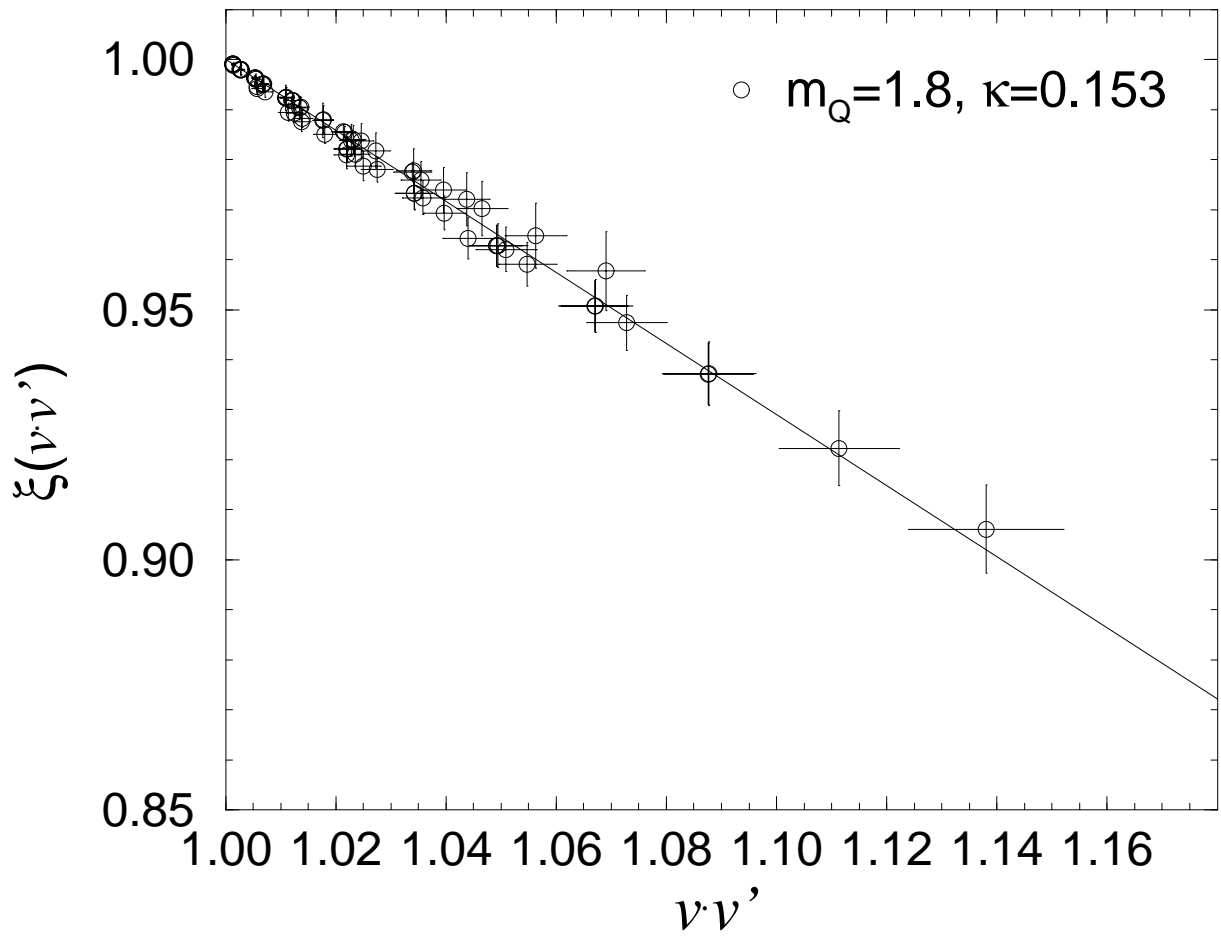
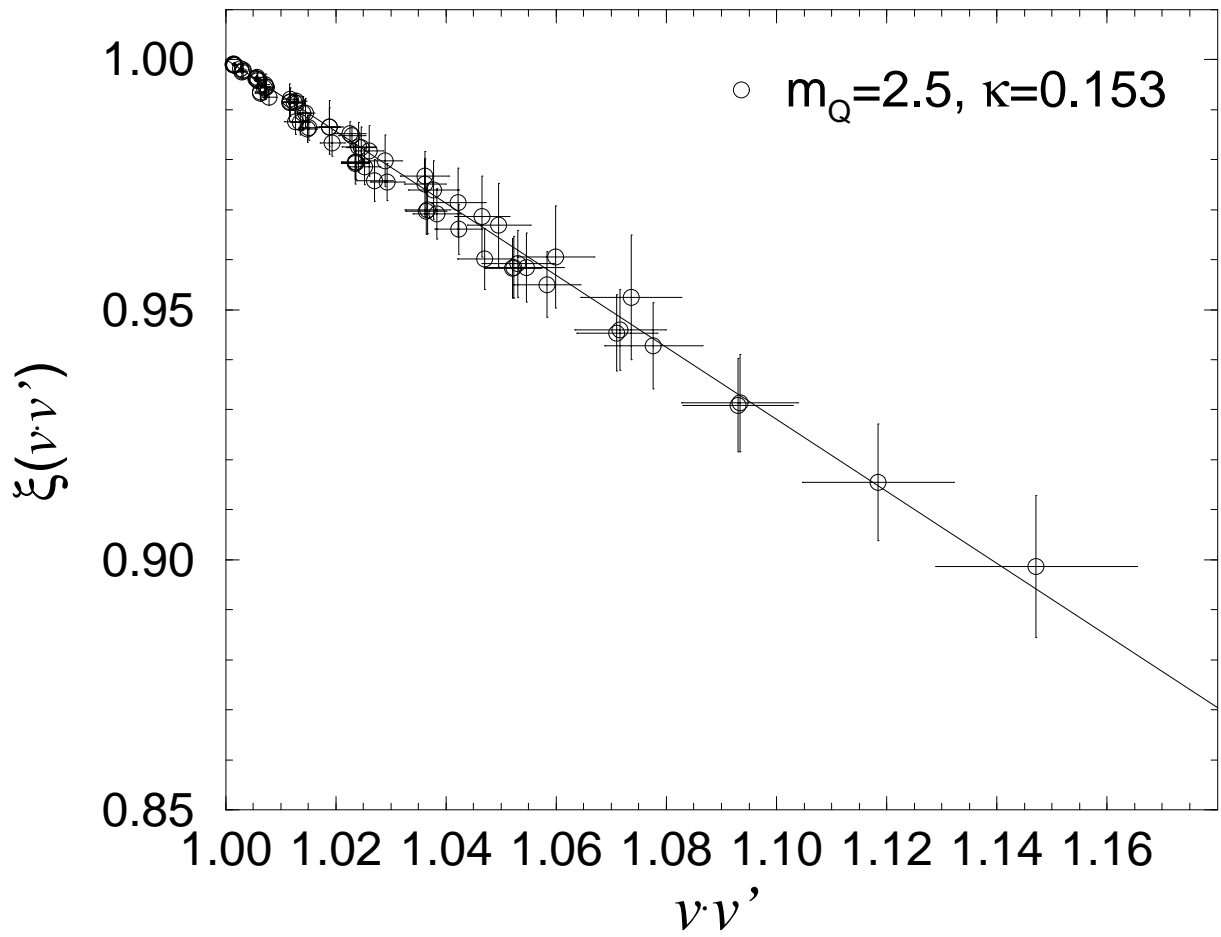
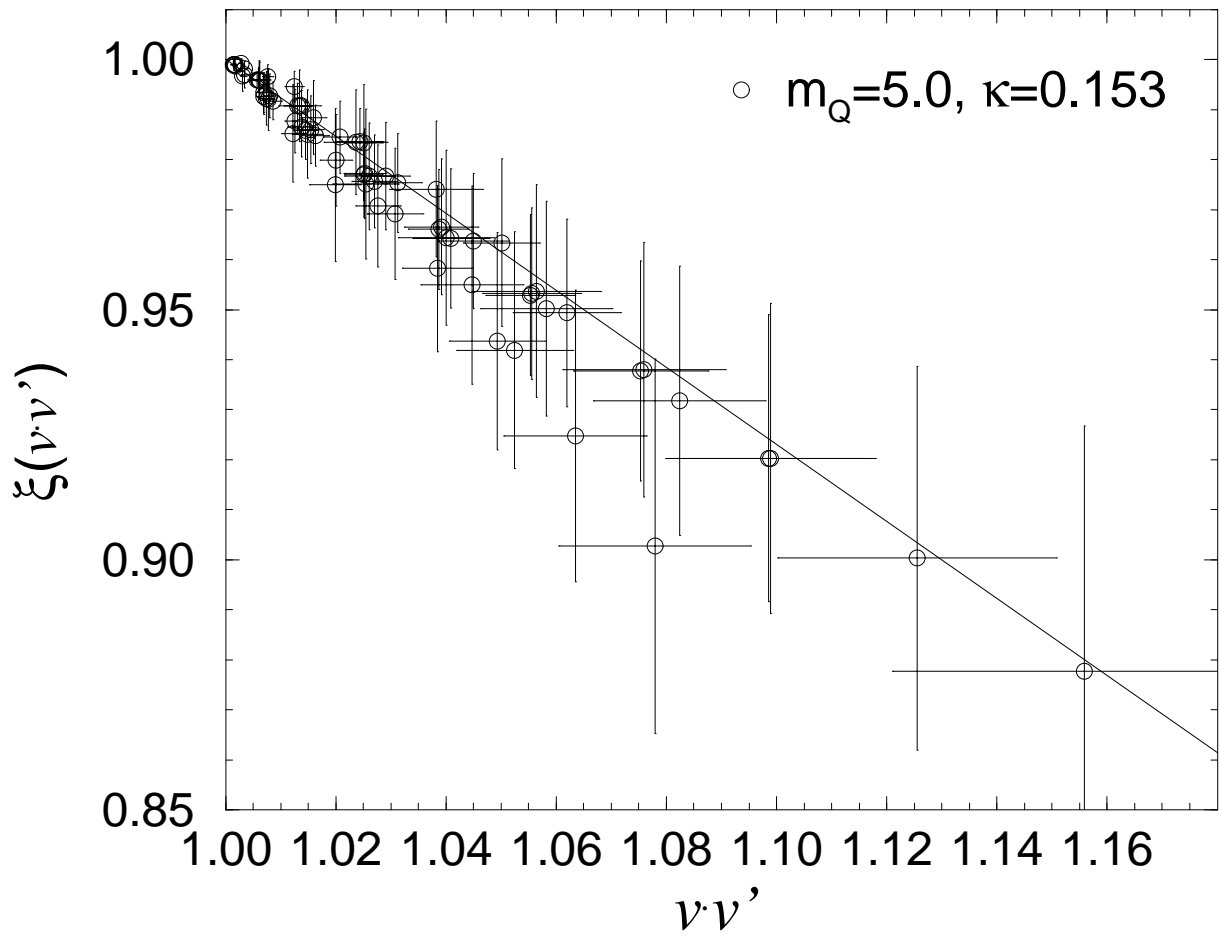
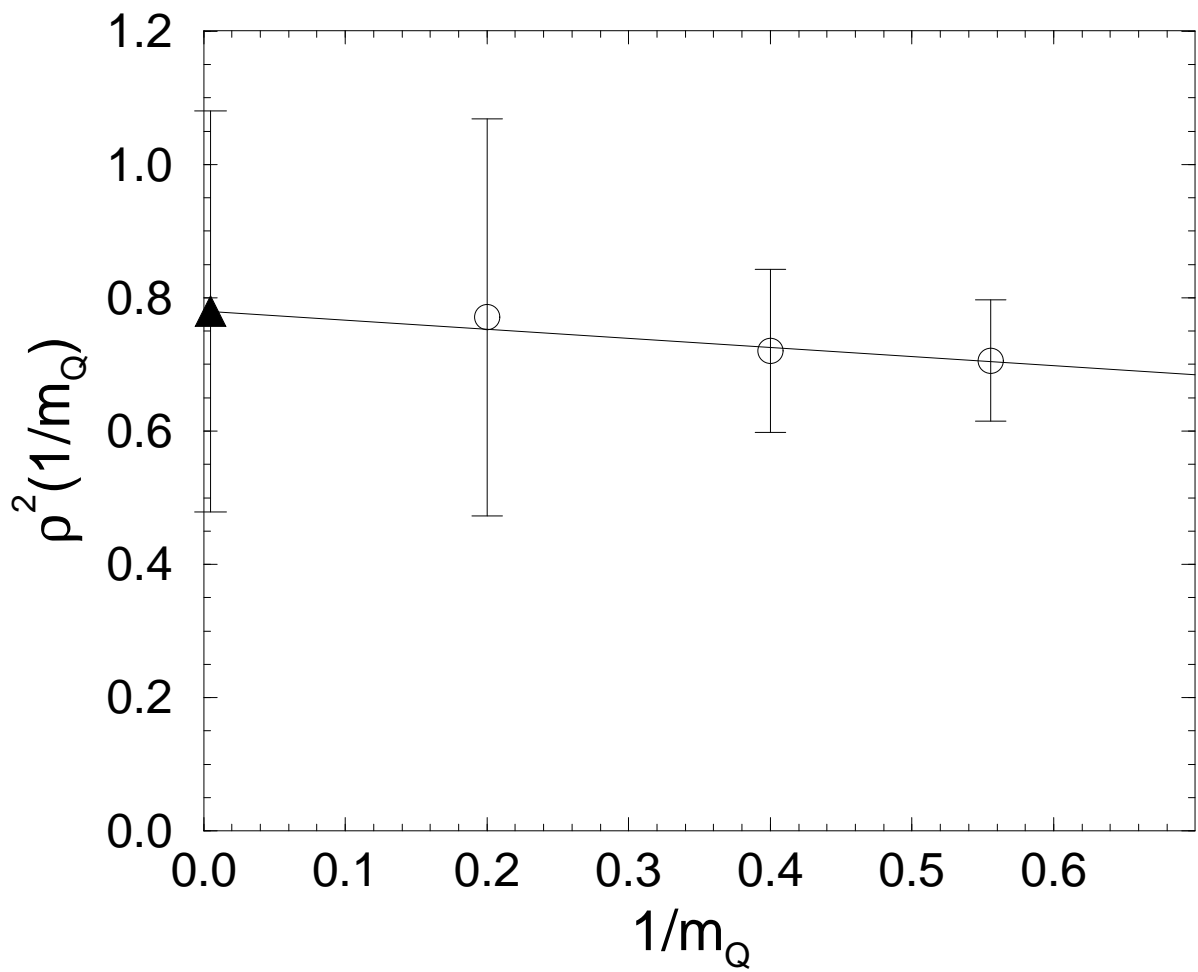
\end{document}